\documentclass[12pt,twocolumn]{article}
\usepackage[a4paper,left=20mm,right=20mm,top=25mm,bottom=25mm,includeheadfoot]{geometry}

\usepackage{mathpazo}
\usepackage{graphicx}
\usepackage{amsmath,textcomp}
\usepackage{makeidx}
\usepackage{multirow}
\makeindex

\usepackage{sectsty}
	\allsectionsfont{\sffamily\raggedright}
	\sectionfont{\sffamily\large\raggedright}
	\subsectionfont{\sffamily\normalsize\raggedright}
\usepackage{ftnright}

\usepackage{widetext}
\usepackage{flushend}
\usepackage{cuted}
\usepackage{color}
\usepackage{graphicx}
\usepackage{hyperref}
\usepackage{pstricks,amssymb}

\usepackage{fancyhdr}
\begin{document}
\title{\vspace{-2em}\bfseries\sffamily Locating Planets in Sky Using 
Manual Calculations}
\author{\normalsize Ashok K. Singal\\[2ex]
Astronomy and Astrophysis Division, Physical Research Laboratory\\
Navrangpura, Ahmedabad 380 009, India.\\
{\tt ashokkumar.singal@gmail.com}
}
\date{\itshape Submitted on xx-xxx-xxxx}
\maketitle

\begin{abstract}
{\sffamily
In this article, we describe a very simple technique to locate naked-eye planets in the sky, to an accuracy of $\sim 1^{\circ}$. The procedure, comprising just three steps, involves very simple manual calculations for planetary orbits around the Sun; all one needs are the initial specifications of planetary positions for some standard epoch and the time periods of their revolutions. After applying a small correction for the orbital ellipticity,   appearance of a planet relative to Sun's position in sky, as seen by an observer from Earth, is found using a scale and a protractor (found inside a school geometry box).
}\\ 
\hrule
\end{abstract}
\section{INTRODUCTION}
Quite often, seeing a bright star-like object in the evening sky (or in the morning sky for the early birds!) many of us would have wondered whether it is a planet and if so, which one. To be able to actually locate a planet in the sky is something that could be thrilling to most of us
and occasionally it provides us an opportunity to impress our friends and acquaintances. 
Although daily planetary positions could be obtained from the professional ephemeris \cite{4} or simply from the internet, yet it is very instructive and much more satisfying to be able to calculate these ourselves, starting from, say, one of Kepler's famous laws, which states that planets go around Sun in elliptical paths.

The path that Earth takes in the sky is called the ecliptic. The familiar Zodiac constellations are just divisions of the ecliptic into twelve parts. Sun as well as planets, as seen from Earth, also appears to move along the ecliptic and pass through the Zodiac constellations. The angle along the ecliptic is called longitude (denoted by $\lambda$), measured eastwards, that is, anti-clockwise, from ${0}^{\circ}$ to ${360}^{\circ}$. Its origin, $\lambda={0}^{\circ}$, is known as the First Point of Aries, the position Sun occupies on the ecliptic around 21st of March every year.  

The first step in our exercise would be to calculate the longitude of a planet as well as that of Earth around Sun. We initially consider the planets to revolve around Sun in uniform circular motion. However, this will entail a correction since the actual orbits are elliptical where the angular speed is not uniform. In second step  we employ a pre-computed table to find the necessary corrections. From another table we also determine distances of planets from Sun. In a third and final step, we plot the distance along the longitude of chosen planet as well as that of Earth centred around Sun on a graph sheet or a chart, using a scale and a protractor (found inside a school geometry box) and measure the angle between the planet and Sun as seen from Earth, which allows us to locate the planet in the sky.

\section{Step 1: Longitude in a circular orbit}

First we consider the planets to move around Sun in circular orbits with uniform angular speeds. We need longitudes of planets in their orbits around Sun on some initial date. Here we take 1st January, 2000, 00:00 UT as our initial date \cite{1} for which we have listed the longitudes ($\lambda_{i}$) of the planets (Table A1).  Also listed in Table A1 is the period $T$  (days) of revolution of each planet \cite{2}. From  $T$ we get the mean angular speed of the planet as
$\omega_{0}=360/T$ ($^{\circ}$/day). We denote the Mean Longitude of the planets 
in the imaginary circular orbit for subsequent dates as $\lambda_{0}$.

As an example, we calculate the mean longitude $\lambda_{0}$ for Venus on 1st October 2018. 

The initial longitude of Venus (on 01.01.2000), $\lambda_{i} = 181.2^{\circ}$.

The mean angular speed of Venus, $\omega_{0} = 1.60213^{\circ}$/day

No. of days between 01.01.2000 and 01.10.2018, $N = 365 \times 18 + 273 + 5 = 6848$ (including 5 leap days).

The mean angle traversed duration this period,  $\omega_{0} N = 1.60213 \times 6848= 10971.4^{\circ}$.

So, on 01.10.18 the mean longitude of Venus, $\lambda_{0} =181.2 + 10971.4  =  11152.6^{\circ}$.

After taking out 30 complete orbits in integer multiple of $360^{\circ}$, we get mean longitude of Venus,
$\lambda_{0}$ $= 11152.6 - 360 \times 30 = 352.6^{\circ}$.

As another example we also calculate the mean longitude of Jupiter on 01.10.18,  $\lambda_{0}=34.3+0.08309 \times 6848=603.3^{\circ}$.
After taking out one complete cycle, the mean longitude of  Jupiter on 01.10.18 is $\lambda_{0}=243.3^{\circ}$.

We also need to calculate the mean longitude of Earth on 01.10.18 as $\lambda_{0}= 100+0.98561 \times 6848=6849.5^{\circ}$.
Or the mean longitude of Earth on 01.10.18 is $\lambda_{0}=9.5^{\circ}$ (after taking out 19 complete orbits).
\section{Step 2: Correcting for the elliptical motion}
The orbits of planets around Sun are elliptical and as a result their angular speeds are not exactly uniform. The corrections in the longitudes of some of the planets due to this variation in their angular speeds could be appreciable. To correct for the elliptical motion, here we employ Table A2, computed based on the formulation derived in \cite{3}. Entries for each planet in Table A2 also account for the orientation 
of its elliptical orbit within the ecliptic plane, specified by the longitude of the perihelion, where perihelion is the point closest to the Sun on the elliptical orbit of the planet.

The correction from the Table A2 for Venus for $\lambda_{0} = 352.6$ is $-0.3^{\circ}$. 
Therefore corrected longitude is $\lambda = 352.6 - 0.3 = 352.3^{\circ}$.

The correction for Jupiter for $=243.3^{\circ}$ is $-3.8^{\circ}$,  
with corrected longitude $\lambda = 243.3 - 3.8 = 239.5^{\circ}$.

Similarly we calculate the corrected elliptical longitude for Earth as $\lambda =7.8^{\circ}$. 
\section{Step 3: Calculating the elongation from Sun}
The difference between the geocentric position of a planet and Sun is called the 
elongation ($\psi$) of the planet and it tells us about planet's position in the sky relative to that of the Sun. As we want to find the sky position of a planet, as seen by an observer located on Earth, we need to find the position of Earth too in the ecliptic. Since Earth longitude changes by $\sim 1^{\circ}$ per day, we need Earth position for the same date and time as that of the planet we are interested in.

We have already calculated the ecliptic longitudes of Venus, Jupiter and Earth. For each of these planets we also need to find radii $r$, the distance from Sun, which is listed against $\lambda_{0}$ in Table A3 in AU (Astronomical Unit - the mean distance between Earth and Sun). For our chosen date of 01/10/2018, $r$ is 0.73 A.U. for Venus, 5.37 for Jupiter and 1.00 A.U. for Earth.

Although one could employ a calculator to compute geocentric longitude and the elongation of the planet \cite{3}, however, a manual geometric construction could be much more illuminating. All one needs is a scale and a protractor, usually found in a school geometry box.
\begin{figure*}[ht]
\includegraphics[width=\linewidth]{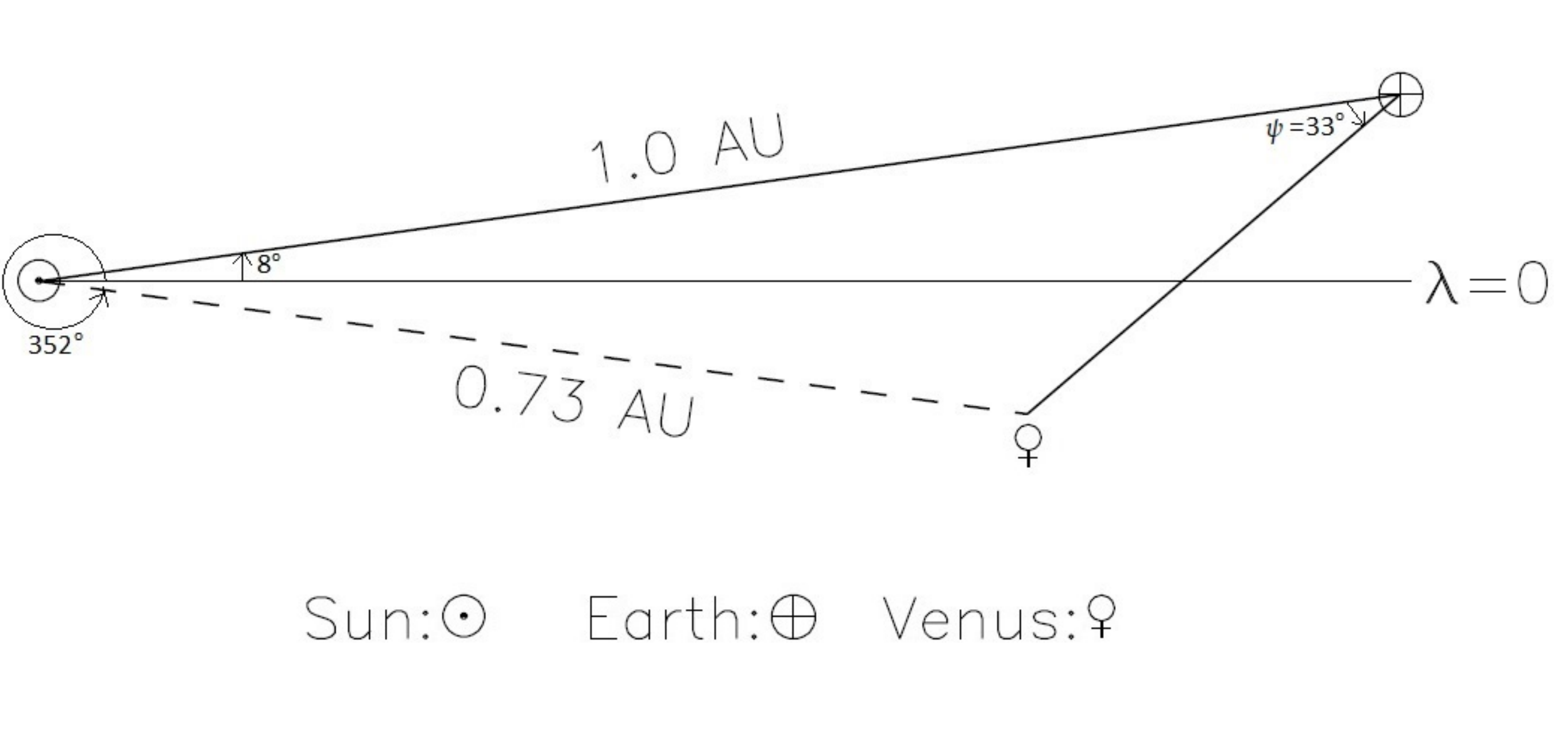}
\caption{Elongation of Venus on 1-10-2018.}
\end{figure*}

Now we plot on a graph sheet or a chart paper the position of Venus at its respective distance 0.73 A.U. (on a suitable scale chosen for 1 A.U.) along its corrected longitude $\lambda=352.3^{\circ}$ (increasing anti-clockwise) around Sun. Similarly we also need to plot the position of Earth on this diagram (see Fig. 1). 

To locate a planet in the sky we determine its elongation $\psi$ which is the angular distance measured eastward (that is anti-clockwise) from Sun`s position on the ecliptic, as seen from Earth (Fig. 1).

That is, from the chart we determine the angle $\psi$ between the line joining Earth to the planet and that from Earth to Sun. If $\psi>0$ (measured anti-clockwise from Earth-Sun line) then the planet position lies to the east of Sun. That means that it will set after the Sun and the planet will be visible above the western horizon in the evening sky. On the other hand, if $\psi<0$, then the planet will rise before the Sun and will be visible in the morning sky above the eastern horizon.

From Fig. 1 we find that on 1/10/2018 Venus is $\sim 33^{\circ}$ east of Sun, and it will be visible at sunset time about $33^{\circ}$ away from Sun's position in the western sky.
\begin{figure*}[]
\includegraphics[width=\linewidth]{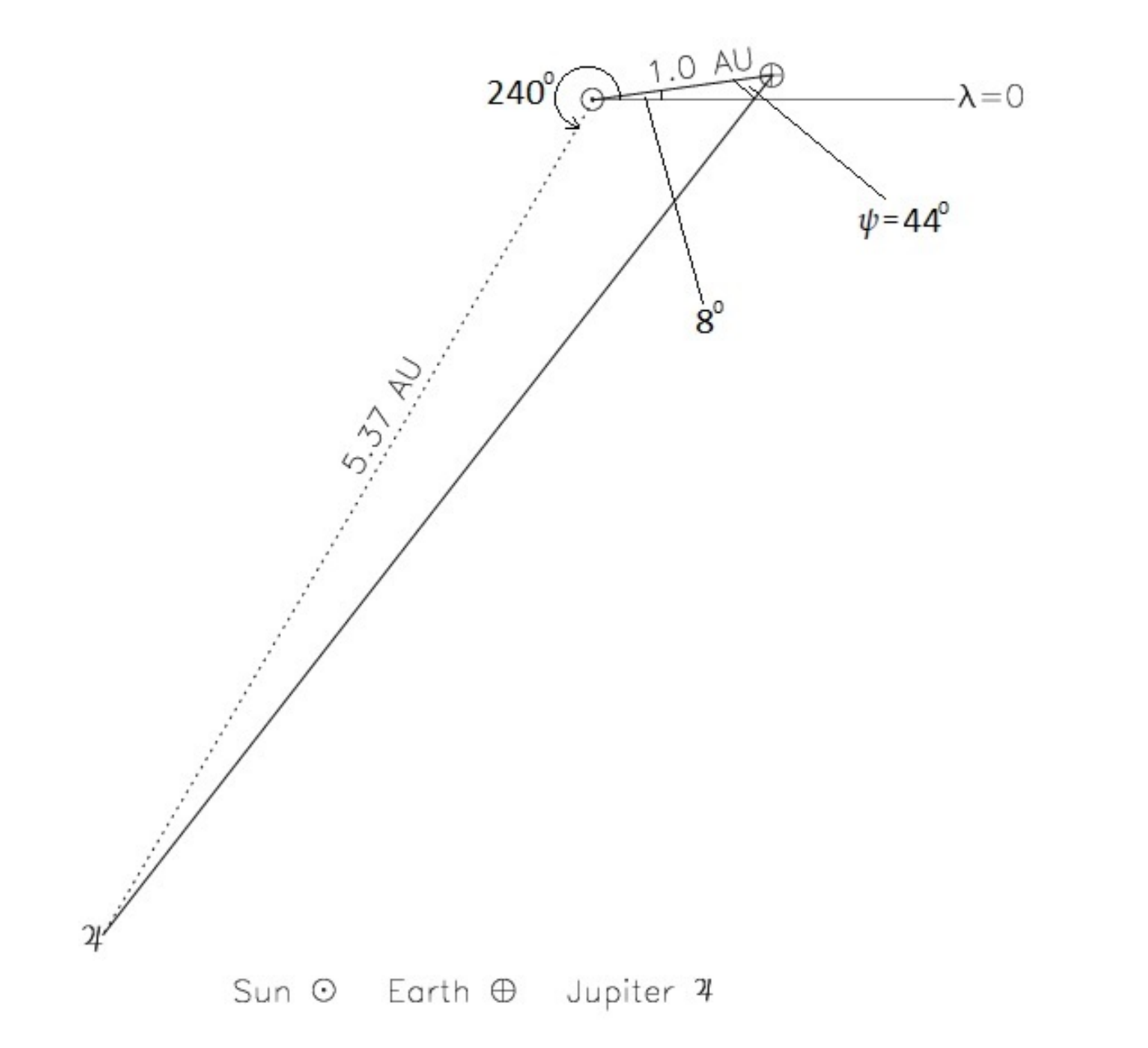}
\caption{{Elongation of Jupiter on 1-10-2018.}}
\end{figure*}

From Fig. 2 we see that on 1/10/2018 Jupiter is $\sim 44^{\circ}$ east of Sun, and it will be therefore visible in the evening about $44^{\circ}$ away from Sun's position. Thus in the evening of 1/10/2018, one will see two bright objects (Venus and Jupiter) separated about $11^{\circ}$ in the western sky.

As Earth completes a full rotation in 24 hours, the westward motion of the sky is at a rate $15^{\circ}$/ hour.
This rate is strictly true for the celestial equator, but we can use this as an approximate rotation rate 
even for the ecliptic, which is inclined at a $23.5^{\circ}$ to the equator. Therefore Venus with an eastern elongation 
$\sim 33^{\circ}$, will be setting a little more than two hours after sunset,  
while Jupiter at $\sim 44^{\circ}$ will set slightly less than three hours after sunset.

Elongations of other planets can be calculated in the same way. In this way, one can easily locate the planets in the sky by knowing their elongations with respect to the Sun.

We may add here that all our calculations have so far been for 00:00 hr UT (Universal time) which corresponds to 05:30 hr IST (Indian Standard Time). However it is possible to determine the planetary positions for any other time of the day. For example for our cases of Venus and Jupiter, which are visible in the evening hours on our chosen date of 1/10/2018, it might be preferable to take number of days in Step 1 as 6848.5, which would then imply that our calculated positions will be for the same date but for 12:00 UT, corresponding to 17:30 hr IST, locally an evening time. 
\begin{table}[t]
\begin{center}
\caption{positions of planets on 01.10.2018 at 
00:00 hr and 12:00 hr UT
}
\vspace{0.3cm}
\vbox{\columnwidth=33pc
\begin{tabular}{lrrcrr}
\hline
\multirow{2}{*}{Planet} 
      & \multicolumn{2}{c}{00:00 hr UT} 
          & \multicolumn{3}{c}{12:00 hr UT} \\            
           \cline{2-3} \cline{5-6}
            & $\lambda$($^{\circ}$)& $\psi$($^{\circ}$) && $\lambda$($^{\circ}$)&  $\psi$($^{\circ}$)  \\
 \hline
Earth & 7.8 & & &  8.3 & \\
Mercury &  213.4  & 7.7 && 214.9 & 8.0   \\
Venus & 352.3 & 32.8 && 353.1 & 32.4\\
Mars &  345.6 & 117.9 && 345.9& 117.6 \\
Jupiter & 239.5 &  44.3 && 239.6 & 43.9 \\
Saturn &  278.9 & 85.4 && 278.9 & 84.9  \\
\hline
\end{tabular}
}
\end{center}
\end{table}

In Table 1, we have listed the elongations of all five naked-eye planets on 01-10-2018 at 00:00 hr UT (5:30 IST) and at 12:00 hr UT (17:30 IST). It should be noted that not only the longitude of each planet around Sun changes by a certain amount, even the longitude of Earth advances by $\sim 1^{\circ}$ in a day, thus affecting the elongation of even Jupiter and Saturn (Table 1), whose angular speeds are relatively small (Table A1).
\section{CONCLUSIONS}
We tried to dispel a general notion that to be able to determine positions of planets in the night sky one requires complex scientific computations, using fast computers. The motive 
of this article has been to impress upon the reader that such accurate calculations are not really necessary for locating naked-eye planets in sky. It was demonstrated  that in just three steps, one can find the positions of planets manually with simple arithmetic calculations. All one needs are the initial specifications of planetary positions for some standard epoch and the time periods of their revolutions. Then after applying a small correction for the orbital ellipticity, the location of a planet in sky, from the point of view of an Earth-based observer, could be found 
and thus one could 
get the thrill of locating a planet at the predicted position in the night sky.

\appendix
\section*{Appendix}
\setcounter{table}{0}
\renewcommand\thetable{{A}\arabic{table}}
\begin{table}[ht]
\begin{center}
\caption{Parameters of the planetary orbits on 01.01.2000, 00:00 UT
}
\vspace{0.3cm}
\vbox{\columnwidth=33pc
\begin{tabular}{lcccccc}
\hline
Planet  & $\lambda_{i}$ ($^{\circ}$) &  $T$ (days) &  $\omega_{0}$ ($^{\circ}$/day)  \\ \hline
Mercury & 250.2 & 87.969 & 4.09235  \\
Venus & 181.2 & 224.701 & 1.60213 \\
Earth & 100.0 & 365.256 & 0.98561  \\ 
Mars & 355.2 & 686.980 & 0.52403  \\
Jupiter & 34.3 & 4332.59 & 0.08309 \\
Saturn & 50.1 & 10759.2 & 0.03346 \\
\hline
\end{tabular}
}
\end{center}
\end{table}
\begin{table*}[ht]
\begin{center}
\vspace{-0.75cm}
\caption{Correction to the longitude for elliptical orbits
}
\vbox{\columnwidth=33pc
\begin{tabular}{rrrrrrr}
\hline
$\lambda_0$ & Mercury  & Venus & Earth & Mars & Jupiter & Saturn \\ 
\hline
   0 &  -24.1 &   -0.4 &   -1.7 &    5.0 &   -1.3 &   -6.1 \\
   10 &  -23.7 &   -0.5 &   -1.7 &    6.7 &   -0.2 &   -6.2 \\
   20 &  -22.4 &   -0.5 &   -1.7 &    8.2 &    0.8 &   -6.0 \\
   30 &  -20.2 &   -0.6 &   -1.6 &    9.4 &    1.8 &   -5.6 \\
   40 &  -17.1 &   -0.6 &   -1.5 &   10.3 &    2.7 &   -5.1 \\
   50 &  -13.2 &   -0.6 &   -1.3 &   10.8 &    3.6 &   -4.4 \\
   60 &   -8.6 &   -0.5 &   -1.1 &   11.0 &    4.3 &   -3.5 \\
   70 &   -3.7 &   -0.5 &   -0.9 &   10.8 &    4.9 &   -2.5 \\
   80 &    1.5 &   -0.4 &   -0.6 &   10.3 &    5.4 &   -1.3 \\
   90 &    6.6 &   -0.3 &   -0.2 &    9.5 &    5.7 &   -0.2 \\
  100 &   11.4 &   -0.2 &    0.1 &    8.5 &    5.8 &    1.0 \\
  110 &   15.6 &   -0.1 &    0.4 &    7.3 &    5.7 &    2.2 \\
  120 &   19.1 &    0.0 &    0.8 &    5.9 &    5.5 &    3.3 \\
  130 &   21.8 &    0.2 &    1.1 &    4.4 &    5.1 &    4.2 \\
  140 &   23.6 &    0.3 &    1.4 &    2.8 &    4.6 &    5.1 \\
  150 &   24.4 &    0.4 &    1.6 &    1.2 &    3.9 &    5.7 \\
  160 &   24.3 &    0.6 &    1.8 &   -0.4 &    3.2 &    6.2 \\
  170 &   23.5 &    0.7 &    2.0 &   -2.1 &    2.4 &    6.5 \\
  180 &   21.9 &    0.8 &    2.1 &   -3.7 &    1.5 &    6.6 \\
  190 &   19.8 &    0.9 &    2.1 &   -5.2 &    0.6 &    6.5 \\
  200 &   17.3 &    0.9 &    2.1 &   -6.6 &   -0.3 &    6.2 \\
  210 &   14.6 &    1.0 &    2.0 &   -7.9 &   -1.2 &    5.7 \\
  220 &   11.6 &    1.0 &    1.9 &   -8.9 &   -2.1 &    5.1 \\
  230 &    8.6 &    1.0 &    1.7 &   -9.8 &   -2.9 &    4.3 \\
  240 &    5.5 &    0.9 &    1.5 &  -10.3 &   -3.6 &    3.5 \\
  250 &    2.5 &    0.9 &    1.2 &  -10.6 &   -4.2 &    2.5 \\
  260 &   -0.6 &    0.8 &    0.9 &  -10.5 &   -4.7 &    1.5 \\
  270 &   -3.6 &    0.7 &    0.6 &  -10.0 &   -5.1 &    0.5 \\
  280 &   -6.7 &    0.6 &    0.3 &   -9.3 &   -5.3 &   -0.5 \\
  290 &   -9.7 &    0.5 &   -0.0 &   -8.1 &   -5.4 &   -1.5 \\
  300 &  -12.7 &    0.4 &   -0.4 &   -6.7 &   -5.2 &   -2.5 \\
  310 &  -15.6 &    0.2 &   -0.7 &   -5.0 &   -4.9 &   -3.4 \\
  320 &  -18.2 &    0.1 &   -0.9 &   -3.1 &   -4.5 &   -4.2 \\
  330 &  -20.5 &   -0.0 &   -1.2 &   -1.1 &   -3.8 &   -4.9 \\
  340 &  -22.4 &   -0.2 &   -1.4 &    1.0 &   -3.1 &   -5.5 \\
  350 &  -23.6 &   -0.3 &   -1.5 &    3.1 &   -2.2 &   -5.9 \\
 \hline
 \end{tabular}
}
\end{center}
\end{table*}
\begin{table*}[ht]
\begin{center}
\vspace{-0.75cm}
\caption{Planetary distances (in AU) corresponding to $\lambda_0$ values
}
\vbox{\columnwidth=33pc
\begin{tabular}{rrrrrrr}
\hline
$\lambda_0$ & Mercury  & Venus & Earth & Mars & Jupiter & Saturn \\ 
\hline
   0 &   0.39 &   0.73 &   1.00 &   1.39 &   4.96 &   9.61 \\
   10 &   0.37 &   0.73 &   1.00 &   1.41 &   4.95 &   9.52 \\
   20 &   0.36 &   0.72 &   1.00 &   1.42 &   4.95 &   9.42 \\
   30 &   0.34 &   0.72 &   1.00 &   1.45 &   4.96 &   9.33 \\
   40 &   0.33 &   0.72 &   0.99 &   1.47 &   4.98 &   9.25 \\
   50 &   0.32 &   0.72 &   0.99 &   1.49 &   5.00 &   9.18 \\
   60 &   0.31 &   0.72 &   0.99 &   1.52 &   5.03 &   9.12 \\
   70 &   0.31 &   0.72 &   0.99 &   1.54 &   5.07 &   9.07 \\
   80 &   0.31 &   0.72 &   0.98 &   1.57 &   5.11 &   9.04 \\
   90 &   0.31 &   0.72 &   0.98 &   1.59 &   5.15 &   9.02 \\
  100 &   0.32 &   0.72 &   0.98 &   1.61 &   5.19 &   9.02 \\
  110 &   0.33 &   0.72 &   0.98 &   1.62 &   5.24 &   9.05 \\
  120 &   0.34 &   0.72 &   0.98 &   1.64 &   5.28 &   9.08 \\
  130 &   0.35 &   0.72 &   0.99 &   1.65 &   5.32 &   9.14 \\
  140 &   0.37 &   0.72 &   0.99 &   1.66 &   5.35 &   9.20 \\
  150 &   0.38 &   0.72 &   0.99 &   1.66 &   5.39 &   9.28 \\
  160 &   0.39 &   0.72 &   0.99 &   1.66 &   5.41 &   9.37 \\
  170 &   0.41 &   0.72 &   0.99 &   1.66 &   5.43 &   9.46 \\
  180 &   0.42 &   0.72 &   1.00 &   1.65 &   5.45 &   9.55 \\
  190 &   0.43 &   0.72 &   1.00 &   1.64 &   5.45 &   9.64 \\
  200 &   0.44 &   0.72 &   1.00 &   1.63 &   5.45 &   9.73 \\
  210 &   0.45 &   0.72 &   1.01 &   1.61 &   5.44 &   9.81 \\
  220 &   0.45 &   0.72 &   1.01 &   1.59 &   5.43 &   9.89 \\
  230 &   0.46 &   0.72 &   1.01 &   1.57 &   5.41 &   9.95 \\
  240 &   0.46 &   0.72 &   1.01 &   1.55 &   5.38 &  10.00 \\
  250 &   0.47 &   0.73 &   1.01 &   1.52 &   5.35 &  10.04 \\
  260 &   0.47 &   0.73 &   1.02 &   1.50 &   5.31 &  10.07 \\
  270 &   0.47 &   0.73 &   1.02 &   1.47 &   5.27 &  10.08 \\
  280 &   0.46 &   0.73 &   1.02 &   1.45 &   5.23 &  10.08 \\
  290 &   0.46 &   0.73 &   1.02 &   1.43 &   5.19 &  10.06 \\
  300 &   0.45 &   0.73 &   1.02 &   1.41 &   5.14 &  10.03 \\
  310 &   0.44 &   0.73 &   1.01 &   1.40 &   5.10 &   9.98 \\
  320 &   0.43 &   0.73 &   1.01 &   1.38 &   5.06 &   9.93 \\
  330 &   0.42 &   0.73 &   1.01 &   1.38 &   5.03 &   9.86 \\
  340 &   0.41 &   0.73 &   1.01 &   1.38 &   5.00 &   9.78 \\
  350 &   0.40 &   0.73 &   1.01 &   1.38 &   4.97 &   9.70 \\
 \hline
 \end{tabular}
}
\end{center}
\end{table*}


\begin{thebibliography}{20}
\bibitem{4}``The Indian Astronomical Ephemeris for the year 2018'', The Indian Meteorological Department, Kolkata (2018)
\bibitem{1} Fr\"{a}nz M. and Harper D.,  Planetary and Space Science 50, 217 (2002)
\bibitem{2} Nicholson I.,  ``Unfolding Our Universe'',  Cambridge University Press (1999)
\bibitem{3} Singal T. and Singal A. K., Prayas, 3, 176 (2008);  arxiv:0910.2778v1
\end{thebibliography}
\end{document}